\documentstyle[11pt,epsfig,amsfonts]{article}
\begin{document}
\title{On two-dimensional superpotentials: \\ from classical Hamilton-Jacobi theory \\
to 2D supersymmetric quantum mechanics}
\author{A. Alonso Izquierdo$^{(a)}$,
M.A. Gonzalez Leon$^{(a)}$ \\ M. de la Torre Mayado$^{(b)}$ and J.
Mateos Guilarte$^{(b)}$
\\ {\normalsize {\it $^{(a)}$ Departamento de Matematica
Aplicada}, {\it Universidad de Salamanca, SPAIN}}\\{\normalsize
{\it $^{(b)}$ Departamento de Fisica} ,{\it Universidad de
Salamanca, SPAIN}}}

\date{}
\maketitle
\begin{abstract}
Superpotentials in ${\cal N}=2$ supersymmetric classical mechanics
are no more than the Hamilton characteristic function of the
Hamilton-Jacobi theory for the associated purely bosonic dynamical
system. Modulo a global sign, there are several superpotentials
ruling Hamilton-Jacobi separable supersymmetric systems, with a
number of degrees of freedom greater than one. Here, we explore
how supersymmetry and separability are entangled in the quantum
version of this kind of system. We also show that the planar
anisotropic harmonic oscillator and the two-Newtonian centers of
force problem admit two non-equivalent supersymmetric extensions
with different ground states and Yukawa couplings.
\end{abstract}

\section{Introduction}
Supersymmetric quantum mechanics was tailor-designed for the
purpose of studying the subtle and crucial concept of spontaneous
supersymmetry breaking by E. Witten \cite{Witten} in a context as
basic and simple as possible. Very soon, the strength of that idea
exploded in an unexpected direction: SUSY quantum mechanics on
N-dimensional Riemannian manifolds \cite{Witten1} provided a
physicist's approach to the very deep index theory of elliptic
operators, with far reaching consequences for the exchange between
the communities of mathematicians and physicists. The physics of
supersymmetric quantum mechanics, however, was mainly studied in
the case of only one degree of freedom. This task proved to be
interesting enough to produce a huge body of literature; here we
quote only References \cite{Casal}, \cite{Ber}, \cite{Khare}, and
\cite{Junker} as the background to our work.

Following previous work on the factorization method on
$N$-dimensional quantum mechanical systems \cite{Andr}, the
general formalism of multi-dimensional supersymmetric quantum
mechanics was established in the mid-eighties by a
Sankt-Petersburg group; see \cite{Andr1}. More recently,
researchers in the entourage of the same group have explored the
interplay between two-dimensional supersymmetric quantum mechanics
with integrability and separability at the classical limit,
\cite{Andr2}, \cite{Can}. In Reference \cite{Aai} we addressed
this problem in a systematic way; we limited ourselves, however,
to the classical theory as our scenario, profiting from the
Hamilton-Jacobi equation to obtain the supersymmetric extension
of classical invariants of Hamilton-Jacobi separable 2D systems.
In the present work, our goal is to address the same issue in a
purely quantum setting. We shall describe how the spectra of
matrix differential operators of different rank are intertwined.
We shall also show that the ground states (zero modes) have a
particularly simple form in this kind of system.

The organization of the paper is as follows: in Section \S 2, for
convenience of the reader, we summarize the general formalism of
${\cal N}=2$ supersymmetric quantum mechanics for systems with $N$
degrees of freedom. In order to set the stage for novel
developments, we briefly rework the theoretical basis of
N-dimensional SUSY quantum mechanics as originally presented in
the papers \cite{Andr}-\cite{Andr1}. They use the Clifford
algebra formalism for the first time in this context, better than
the exterior calculus of \cite{Witten} for our purposes. We also
try to adapt this framework to the cohomological approach
proposed in Reference \cite{Kir} to solve the supersymmetric
Coulomb problem in any dimension by algebraic means. The
entanglement between Hamilton-Jacobi theory and the separation of
variables of the quantum Schr\"odinger equation is examined in
Section \S 3. In Section \S 4 we discuss two interesting
two-dimensional physical systems. Finally, we offer a brief
Summary in Section \S 5.

\section{${\cal N}=2$ supersymmetric quantum mechanics}
\subsection{$N$-dimensional ${\cal N}=2$ SUSY quantum mechanics}
Let $\gamma^j$, $\gamma^{N+j}$, $j=1, 2, \cdots , N$ be the
Hermitian generators, $(\gamma^j)^\dagger=\gamma^j$,
$(\gamma^{N+j})^\dagger=\gamma^{N+j}$, of the Clifford algebra
${\bf C}({\bf R}^{2N})$ of ${\bf R}^{2N}$: $
\{\gamma^j,\gamma^k\}= 2 \delta^{jk}$,
$\{\gamma^{N+j},\gamma^{N+k}\}=2 \delta^{jk}$,
$\{\gamma^j,\gamma^{N+k}\}=0$, where $\{ .  , .\}$ denotes
anticommutator. Because the dimension of the irreducible
representation of ${\bf C}({\bf R}^{2N})$ is $\sum_{f=0}^N  {N
\choose f } =2^N$, the generators of ${\bf C}({\bf R}^{2N})$ are
$2^N\times 2^N$ Hermitian matrices. The linear combinations
$\psi^j_+=\frac{1}{2}(\gamma^j-i\gamma^{N+j})\, , \,
\psi^j_-=\frac{1}{2}(\gamma^j+i\gamma^{N+j}) $ of the generators
satisfy the anticommutation rules: $\{\psi^j_+ , \psi^k_-
\}=\delta^{jk}$, $ \{\psi^j_+ , \psi^k_+ \}=\{\psi^j_- , \psi^k_-
\}=0$. Thus, $\psi^j_+$ and $\psi^j_-$ can be thought of as \lq\lq
creation" and \lq \lq annihilation" fermionic operators. From
these operators one defines the fermionic total number operator,
$\hat{f}=\sum_{j=1}^N\psi^j_+\psi^j_-$, which allows one to assign
a grading to the space of the irreducible representation of the
Clifford algebra -the fermionic Fock space-: ${\cal F}={\cal
F}_0\oplus {\cal F}_1\oplus \cdots \oplus {\cal F}_N
=\oplus_{f=0}^N \,  {\cal F}_{f} $, $ \hat{f}{\cal F}_{f} =f{\cal
F}_{f}$, $\psi_+: {\cal F}_f\longrightarrow {\cal F}_{f+1}$,
$\psi_-: {\cal F}_f\longrightarrow {\cal F}_{f-1} $.

The key ingredients in defining a $N$-dimensional quantum
mechanical \footnote{We take a system of units where $\hbar=1$.}
system with ${\cal N}=2$ supersymmetry are the supercharges:
\begin{eqnarray}
\hat{Q}_+&=&e^{W(x^1,\cdots ,x^N)}\hat{Q}_+^0e^{- W(x^1,\cdots
,x^N)}= i\sum_{j=1}^N\psi^j_+\left(\frac{\partial}{\partial
x^j}-\frac{\partial W}{\partial x^j}\right) \hspace{0.3cm} ,
\hspace{0.3cm}
\hat{Q}_+^0=i\sum_{j=1}^N\psi^j_+\frac{\partial}{\partial
x^j}\hspace{0.3cm} , \label{eq:scha} \\
\hat{Q}_-&=&e^{-W(x^1,\cdots ,x^N)}\hat{Q}_-^0e^{W(x^1,\cdots
,x^N)} =i\sum_{j=1}^N\psi^j_-\left(\frac{\partial}{\partial
x^j}+\frac{\partial W}{\partial x^j}\right) \hspace{0.3cm} ,
\hspace{0.3cm}
\hat{Q}_-^0=i\sum_{j=1}^N\psi^j_-\frac{\partial}{\partial x^j}
\hspace{0.3cm} , \label{eq:schaa}
\end{eqnarray}
which change the number of fermions, $ \hat{Q}_+: {\cal
F}_f\longrightarrow {\cal F}_{f+1} $, $ \hat{Q}_-: {\cal
F}_f\longrightarrow {\cal F}_{f-1}$, and close the ${\cal N}=2$
SUSY algebra:
\begin{equation}
\{\hat{Q}_+ , \hat{Q}_-\}=2\hat{H} \hspace{0.4cm},\hspace{0.4cm}
[\hat{Q}_+ , \hat{H}] = [\hat{Q}_- , \hat{H}] =0
\hspace{0.4cm},\hspace{0.4cm}\hat{Q}_+^2=0
\hspace{0.4cm},\hspace{0.4cm} \hat{Q}_-^2=0 \qquad .
\label{eq:salg}
\end{equation}
Here, ${\hat H}$ is the ${\hat Q}_\pm$-invariant Hamiltonian:
\begin{equation}
\hat{H}=-\frac{1}{2}\sum_{j=1}^N\left(\frac{\partial}{\partial
x^j}+\frac{\partial W}{\partial
x^j}\right)\left(\frac{\partial}{\partial x^j}-\frac{\partial
W}{\partial x^j}\right){\bf
I}_{2^N}-\sum_{j=1}^N\sum_{k=1}^N\frac{\partial^2 W}{\partial
x^j\partial x^k}\psi^j_+\psi^k_- \qquad . \label{eq:sham}
\end{equation}
Compare these expressions with the Hamiltonians, supercharges and
SUSY algebra of \cite{Andr1} (Section \S 3) and \cite{Kir}
(Section \S 3). The Hilbert space of states ${\cal H}={\cal
F}\otimes L^2({\bf R}^N)$ inherits a grading from the fermionic
Fock space:
\[
{\cal H}={\cal H}_0\oplus {\cal H}_1\oplus\cdots \oplus{\cal
H}_{N-1}\oplus {\cal H}_N=\oplus_{f=0}^N{\cal H}_f \qquad , \qquad
{\cal H}_f={\cal F}_f\otimes L^2({\bf R}^N)\qquad .
\]
Let us choose an orthonormal basis $\vec{e}_j, \, j=1,2, \cdot , N
, \, \vec{e}_j.\vec{e}_k=\delta_{jk}$ in ${\bf R}^N$. The
Hamiltonian acting on ${\cal H}_0$ is an ordinary Schr\"odinger
operator with potential energy:
\begin{equation}
\hat{V}(\vec{x})={1\over 2}\left(
\vec{\nabla}W(\vec{x})\vec{\nabla}W(\vec{x})+
\nabla^2W(\vec{x})\right) \qquad , \qquad
\vec{\nabla}=\sum_{j=1}^N{\partial\over\partial x_j}\vec{e}_j
\qquad , \qquad \nabla^2=\sum_{j=1}^N{\partial^2\over\partial
x_j^2} \label{eq:ric}
\end{equation}
i.e. it is obtained from the gradient and the Laplacian of the
function $W$, called the superpotential for this reason. Acting on
${\cal H}_f$, however, $\hat{H}$ is a ${ N \choose f}\times { N
\choose f}$-matrix of differential operators but all the
interactions are also determined by the superpotential $W$, see
(\ref{eq:sham}). In particular, the Yukawa terms -interactions
sensitive to the fermionic number of the state- depend on the
second partial derivatives of $W$. $W$ fully determines the
supersymmetric mechanical system.

There is perfect analogy with de Rahm cohomology, see also
\cite{Andr1} (Section \S 4). The SUSY charges play the r$\hat{\rm
o}$le of the exterior derivative and its adjoint such that, in
the SUSY complex,
\[
{\cal H}_0 {{{{\hat Q}_+}\atop{\longrightarrow}}\atop
{{\longleftarrow}\atop{{\hat Q}_-}}} {\cal H}_1{{{{\hat
Q}_+}\atop{\longrightarrow}}\atop {{\longleftarrow}\atop{{\hat
Q}_-}}}{\cal H}_2{{{{\hat Q}_+}\atop{\longrightarrow}}\atop
{{\longleftarrow}\atop{{\hat Q}_-}}}\cdots\cdots{{{{\hat
Q}_+}\atop{\longrightarrow}}\atop {{\longleftarrow}\atop{{\hat
Q}_-}}}{\cal H}_{N-1}{{{{\hat Q}_+}\atop{\longrightarrow}}\atop
{{\longleftarrow}\atop{{\hat Q}_-}}}{\cal H}_N \qquad ,
\]
one defines the SUSY cohomology groups: $H^f({\cal H},{\bf
C})=\frac{{\rm Ker}\,{\hat Q}_+^f}{{\rm Im}\,{\hat Q}_+^{f-1}}$ \,
\,. Because the supercharges are nilpotent, there is a Hodge-type
decomposition theorem - ${\cal H}={\hat Q}_+{\cal H}\oplus {\hat
Q}_-{\cal H}\oplus {\rm Ker}{\hat H}$ - where the kernel of
$\hat{H}$ is a finite-dimensional subspace spanned by the zero
modes. The proof is easy: invert $\hat{H}$ on the orthogonal
subspace to ${\rm Ker}\hat{H}$ and write:
\[
{\cal H}^{\perp}=\frac{{\hat Q}_+{\hat Q}_-+{\hat Q}_-{\hat
Q}_+}{{\hat H}}{\cal H}^{\perp} ={\hat Q}_+\left({{\hat Q}_-\over
{\hat H}}{\cal H}^{\perp}\right)+{\hat Q}_-\left({{\hat
Q}_+\over{\hat H}}{\cal H}^{\perp}\right)\qquad .
\]
$\hat{H}$ plays the r$\hat{\rm o}$le of the Laplacian and we talk
about $\hat{Q}_\pm$-exact and $\hat{H}$-harmonic states.

As in Hodge theory, zero modes play a special r$\hat{{\rm o}}$le.
$E=0$ eigenfunctions (zero modes) satisfy
$\hat{Q}_+\Psi_0^f=\hat{Q}_-\Psi_0^f=0$: $\Psi_0^f\in {\rm
Ker}\hat{H}$. If $\Psi_0^f=\hat{Q}_+\Phi_0^{f-1}$,
$\hat{Q}_-\hat{Q}_+\Phi_0^{f-1}=0$ implies that
$||\Psi_0^f||=||\hat{Q}_+\Phi_0^{f-1}||=0$. Thus, non-trivial
zero-energy states are all the $\hat{Q}_\pm$-closed states that
are not $\hat{Q}_\pm$-exact. Spontaneous supersymmetry breaking
will occur if all the cohomology groups $H^f({\cal H},{\bf C})$
are trivial. The Witten index is the Euler characteristic of the
SUSY complex: ${\rm Tr}(-1)^{\hat{f}}=\sum_{f_+}{\rm
dim}H^{f_+}({\cal H},{\bf C})-\sum_{f_-}{\rm dim}H^{f_-}({\cal
H},{\bf C})$, where $f_+(f_-)$ runs over even(odd) numbers of
fermions; see \cite{Witten}. This index is frequently used to
decide whether or not a given system presents supersymmetry
breaking because ${\rm Tr}(-1)^{\hat{f}}$ is easier to compute
than the cohomology groups.

\subsection{Two-dimensional ${\cal N}=2$ SUSY quantum mechanics}
In systems with $N=2$ degrees of freedom, the formalism of ${\cal
N}=2$ supersymmetric quantum mechanics can be developed quite
explicitly. Creation and annihilation fermionic operators are
defined from the four dimensional Dirac/Majorana matrices: {\small
\begin{eqnarray*}
\psi_+^1=\frac{1}{2}(\gamma^1-i\gamma^3)=\left(\begin{array}{cccc}
0 & 0 & 0 & 0 \\ 1 & 0 & 0 & 0 \\ 0 & 0 & 0 & 0 \\ 0 & 0 & 1 & 0
\end{array}\right)\quad &,& \psi_-^1=\frac{1}{2}(\gamma^1+i\gamma^3)=\left(\begin{array}{cccc}
0 & 1 & 0 & 0 \\ 0 & 0 & 0 & 0 \\ 0 & 0 & 0 & 1 \\ 0 & 0 & 0 & 0
\end{array}\right)\\
\psi_+^2=\frac{1}{2}(\gamma^2-i\gamma^4)=\left(\begin{array}{cccc}
0 & 0 & 0 & 0 \\ 0 & 0 & 0 & 0 \\ 1 & 0 & 0 & 0 \\ 0 & -1 & 0 & 0
\end{array}\right)\quad &,& \psi_-^2=\frac{1}{2}(\gamma^2+i\gamma^4)=\left(\begin{array}{cccc}
0 & 0 & 1 & 0 \\ 0 & 0 & 0 & -1 \\ 0 & 0 & 0 & 0 \\ 0 & 0 & 0 & 0
\end{array}\right)\qquad ,
\end{eqnarray*}}
which are related to the operators $b_1^\dagger$ and $b_2^\dagger$
defined in \cite{Andr1} (Section \S 5). The supercharges are the
$2\times 2$-matrices of differential operators: {\footnotesize
\[
\hat{Q}_+=i{\small\left(\begin{array}{cccc} 0 & 0 & 0 & 0 \\
{\partial \over \partial x^1}-{\partial W \over \partial x^1} & 0
& 0 & 0 \\ {\partial \over\partial x^2}-{\partial W \over \partial
x^2} & 0 & 0 & 0
\\ 0 & -{\partial \over\partial x^2}+{\partial W \over\partial x^2} & {\partial
\over\partial x^1}-{\partial W \over\partial x^1} & 0
\end{array}\right)}\, ; \, \hat{Q}_-=i{\small\left(\begin{array}{cccc} 0 & {\partial \over \partial
x^1}+{\partial W \over \partial x^1} & {\partial \over \partial
x^2}+{\partial W \over \partial x^2} & 0
\\ 0 & 0 & 0 & -{\partial \over \partial x^2}-{\partial W \over \partial x^2}
\\ 0 & 0 & 0 & {\partial \over \partial x^1}+{\partial W \over \partial x^1}
\\ 0 & 0 & 0 & 0
\end{array}\right)}\qquad ,
\]}
which are nilpotent: $\hat{Q}_+^2=0=\hat{Q}^2_-$. The SUSY algebra
\[
\{\hat{Q}_+ , \hat{Q}_- \}=2\hat{H} \qquad , \qquad
[\hat{Q}_+,\hat{H}]=[\hat{Q}_-,\hat{H}]=0
\]
closes in a Hamiltonian of the form
\[
2\hat{H}=2\left(\begin{array}{ccc} \hat{h}^{(0)} & 0 & 0 \\ 0 &
\hat{h}^{(1)} & 0
\\ 0 & 0 & \hat{h}^{(2)}
\end{array}\right)\qquad ,
\]
where
\begin{equation}
2\hat{h}^{(f=0)}=-\nabla^2+\vec{\nabla}W\vec{\nabla}W+\nabla^2W
\qquad {\rm and} \qquad 2\hat{h}^{(f=2)}=
-\nabla^2+\vec{\nabla}W\vec{\nabla}W-\nabla^2W \qquad ,
\label{eq:bham}
\end{equation}
are ordinary Schr\"odinger operators, and {\small \begin{equation}
2\hat{h}^{(f=1)}=\left(\begin{array}{cc}
-\nabla^2+\vec{\nabla}W\vec{\nabla}W-\Box^2W & -2{\partial^2 W
\over\partial x^1\partial x^2} \\  -2{\partial^2 W \over\partial
x^1\partial x^2} & -\nabla^2+\vec{\nabla}W\vec{\nabla}W+\Box^2W
\end{array}\right)\, ,\, \Box^2={\partial^2\over\partial x^1\partial
x^1}-{\partial^2\over\partial x^2\partial x^2} \qquad ,
\label{eq:fham}
\end{equation}}
is a $2\times 2$- matrix Schr\"odinger operator; see \cite{Andr1},
Section \S 5.

Given an eigenstate of $\hat{H}$ in ${\cal H}_0$ with $E\neq 0$,
\[
\hat{h}^{(0)}\psi_E(x^1,x^2)=E\psi_E(x^1,x^2) \qquad , \qquad
\Psi_E^{(0)}(x^1,x^2)={\footnotesize\left(\begin{array}{c}
\psi_E(x^1,x^2)
\\ 0
\\ 0
\\ 0
\end{array}\right)}\qquad ,
\]
we have that $\hat{Q}_-\Psi_E^{(0)}(x^1,x^2)=0$ -it is
$\hat{Q}_-$-closed-. However, {\small
\[
\hat{Q}_+\Psi_E^{(0)}(x^1,x^2)=i{\small \left(\begin{array}{c} 0
\\ (\frac{\partial}{\partial x^1}- \frac{\partial W}{\partial
x^1})\psi_E(x^1,x^2)
\\ ({\partial\over\partial x^2}-{\partial W\over\partial x^2})\psi_E(x^1,x^2) \\ 0
\end{array}\right)}
\]}
is a eigenstate of $\hat{H}$ with the same energy and fermionic
number $f=1$:
\[
\hat{h}^{(1)}{\small \left(\begin{array}{c}
(\frac{\partial}{\partial x^1}- \frac{\partial W}{\partial
x^1})\psi_E(x^1,x^2)
\\ ({\partial\over\partial x^2}-{\partial W\over\partial x^2})\psi_E(x^1,x^2)
\end{array}\right)}=E{\small \left(\begin{array}{c} (\frac{\partial}{\partial x^1}- \frac{\partial W}{\partial
x^1})\psi_E(x^1,x^2)
\\ ({\partial\over\partial x^2}-{\partial W\over\partial x^2})\psi_E(x^1,x^2)
\end{array}\right)}\qquad ;
\]
$\hat{h}^{(0)}$ is intertwined with $\hat{h}^{(1)}$ and one says
that $\Psi_E^{(1)}=\hat{Q}_+\Psi_E^{(0)}$ is a $\hat{Q}_+$-exact
state. Simili modo, starting from eigenstates of $\hat{H}$ with
$E\neq 0$ in ${\cal H}_2$ -all of them $\hat{Q}_+$-closed, i.e.
$\hat{Q}_+\Psi_E^{(2)}=0$- ,
\[
\hat{h}^{(2)}\phi_E(x^1,x^2)=E\phi_E(x^1,x^2) \qquad , \qquad
\Psi_E^{(2)}(x^1,x^2)={\footnotesize\left(\begin{array}{c} 0 \\ 0 \\ 0 \\
\phi_E(x^1,x^2)
\end{array}\right)}\qquad ,
\]
one easily sees that - the $\hat{Q}_-$-exact state-
\[ \hat{Q}_-\Psi_E^{(2)}(x^1,x^2)=i{\footnotesize
\left(\begin{array}{c} 0
\\ -(\frac{\partial}{\partial x^2}+\frac{\partial W}{\partial
x^2})\phi_E(x^1,x^2)
\\ ({\partial\over\partial x^1}+{\partial W\over\partial x^1})\phi_E(x^1,x^2) \\ 0
\end{array}\right)}
\]
is a eigenstate of $\hat{H}$:
\[
\hat{h}^{(1)}{\small \left(\begin{array}{c}
(-\frac{\partial}{\partial x^2}- \frac{\partial W}{\partial
x^2})\phi_E(x^1,x^2)
\\ ({\partial\over\partial x^1}+{\partial W\over\partial x^1})\phi_E(x^1,x^2)
\end{array}\right)}=E{\small \left(\begin{array}{c} (-\frac{\partial}{\partial x^2}-
\frac{\partial W}{\partial x^2})\phi_E(x^1,x^2)
\\ ({\partial\over\partial x^1}+{\partial W\over\partial x^1})\phi_E(x^1,x^2)
\end{array}\right)} \qquad .
\]
$\hat{h}^{(2)}$ and $\hat{h}^{(1)}$ are also intertwined. Note,
however, that
$\langle\hat{Q}_-\Psi_E^{(2)}|\hat{Q}_+\Psi_E^{(0)}\rangle =0$ and
$\hat{h}^{(0)}$ is not intertwined with $\hat{h}^{(2)}$. See
Reference \cite{Andr3} to find how two scalar Hamiltonians are
intertwined through second-order supercharges.

\subsection{Zero energy eigenstates: spontaneous symmetry
breaking}

The zero energy wave functions for the scalar Hamiltonians satisfy
respectively: $ \hat{Q}_+\Psi_0^{(0)}(x^1,x^2)=0 $,
$\hat{Q}_-\Psi_0^{(2)}(x^1,x^2)=0 $. Therefore, $\vec{\nabla}\log
\psi_0(x^1,x^2)=\vec{\nabla}W$, $\vec{\nabla}\log
\phi_0(x^1,x^2)=-\vec{\nabla}W$, and
\begin{equation}
\Psi_0^{(0)}(x^1,x^2)=C{\footnotesize \left(
\begin{array}{c} {\rm exp}[W(x^1,x^2)]
\\ 0 \\ 0 \\ 0
\end{array} \right)} \qquad , \qquad \Psi_0^{(2)}(x^1,x^2)=C{\footnotesize \left(
\begin{array}{c} 0
\\ 0 \\ 0 \\ {\rm exp}[-W(x^1,x^2)]
\end{array} \right)} \qquad . \label{eq:bzer}
\end{equation}
There are normalizable zero-energy states in ${\cal H}_0$ or
${\cal H}_2$ - and $H^{f=0}({\cal H},{\bf C})$ or $H^{f=2}({\cal
H},{\bf C})$ are non-trivial- if
\[
\int\int_{{\Bbb R}^2}\,dx^1dx^2\, e^{2W(x^1,x^2)}<+\infty \qquad
{\rm or} \qquad \int\int_{{\Bbb R}^2}\,dx^1dx^2\,
e^{-2W(x^1,x^2)}<+\infty \qquad .
\]
Unbroken supersymmetry due to bosonic zero modes arise in 2D SUSY
quantum mechanics under the same requirements as in 1D SUSY
quantum mechanics, see \cite{Cecotti}. However, the search for
wave functions belonging to ${\rm Ker}\, \hat{h}^{(1)}$ is
slightly more difficult.
\[
\hat{Q}_-\Psi_0^{(1)}(x^1,x^2)=0=\hat{Q}_+\Psi_0^{(1)}(x^1,x^2)
\]
requires integration of the equations
\begin{equation}
\vec{\nabla} \log \xi_0(x^1,x^2)= -{\partial W\over\partial
x^1}\vec{e}_1+{\partial W\over\partial x^2}\vec{e}_2 \quad ,
\quad  \vec{\nabla}\log \eta_0(x^1,x^2)= {\partial W\over\partial
x^1}\vec{e}_1-{\partial W\over\partial x^2}\vec{e}_2 \quad .
\label{eq:fzer}
\end{equation}
Note that in the odd cases the gradient of the log of the wave
function is equal to the gradient of the superpotential on a plane
with the reverse orientation. The solutions of (\ref{eq:fzer})
are:
\begin{equation}
\Psi_0^{(1)}(x^1,x^2)=C_1{\footnotesize \left(
\begin{array}{c} 0 \\ {\rm exp}[\tilde{W}(x^1,x^2)]
\\ 0 \\ 0
\end{array} \right)} + C_2{\footnotesize \left(
\begin{array}{c} 0
\\ 0 \\ {\rm exp}[-\tilde{W}(x^1,x^2)] \\ 0
\end{array} \right)}=C_1{\footnotesize \left(\begin{array}{c} 0 \\ \xi_0(x^1,x^2) \\ 0 \\ 0
\end{array}\right)}
+C_2{\footnotesize \left(\begin{array}{c} 0 \\ 0 \\
\eta_0(x^1,x^2) \\ 0 \end{array}\right)} \qquad , \label{eq:bzer1}
\end{equation}
where $\tilde{W}$ is such that: $\frac{\partial\tilde{W}}{\partial
x^1}=-\frac{\partial W}{\partial x^1} \, , \,
\frac{\partial\tilde{W}}{\partial x^2}=\frac{\partial W}{\partial
x^2}$. There are normalizable zero-energy states in ${\cal H}_1$ -
and $H^{f=1}({\cal H},{\bf C})$ is non-trivial- if either
\[
\int\int_{{\Bbb R}^2}\,dx^1dx^2\, e^{2\tilde{W}(x^1,x^2)}<+\infty
\qquad {\rm and}/{\rm or} \qquad \int\int_{{\Bbb R}^2}\,dx^1dx^2\,
e^{-2\tilde{W}(x^1,x^2)}<+\infty \qquad .
\]
There are requirements on the superpotential to find unbroken
supersymmetry coming from fermionic zero modes similar to those
met in the bosonic sectors.

\section{Hamilton-Jacobi theory, supersymmetry and separability}
The quantum system described in Section \S 2 enjoys ${\cal N}=2$
supersymmetry by construction; the datum needed to set the
interactions is the superpotential $W(\vec{x})$. Alternatively,
there might be interest in knowing if a given Hamiltonian admits
${\cal N}=2$ supersymmetry; in that case, the datum is the
potential energy $\hat{V}(\vec{x})$ and the identification of the
superpotential requires that the Riccati-like PDE (\ref{eq:ric})
must be solved. In \cite{Kir}, the superpotential for the quantum
Coulomb problem is shown to be:
$W(x_1,x_2)=\sqrt{2\lambda}\sqrt{x_1^2+x_2^2}$. Temporarily
recovering the Planck constant, one finds:
\[
{1\over 2}\vec{\nabla}W\vec{\nabla}W=\lambda \hspace{0.5cm} ;
\hspace{0.5cm} {1\over 2}\left(\vec{\nabla}W\vec{\nabla}W\pm\hbar
\nabla^2W \right)=\lambda\left[1\pm{\hbar\over 2}
\sqrt{{2\over\lambda}}\cdot{1\over r} \right] \qquad .
\]
The classical and zero-Grasmann limit of this supersymmetric
system is therefore the free particle; the second partial
derivatives of the superpotential arising in $\hat{h}^{(1)}$ are
also multiplied by $\hbar$.

In \cite{Heumann}, the superpotential for the supersymmetric
Coulomb problem is chosen in such a way that the Coulomb potential
energy arises at the classical non-Grassman limit:
$W(x_1,x_2)=2\sqrt{2\lambda}(x_1^2+x_2^2)^{{1\over 4}}$ is the
solution of the Hamilton-Jacobi equation for the Coulomb problem,
instead of (\ref{eq:ric}):
\[
{1\over 2}\vec{\nabla}W\vec{\nabla}W={\lambda\over r}
\hspace{0.5cm} ; \hspace{0.5cm} {1\over
2}\left(\vec{\nabla}W\vec{\nabla}W\pm\hbar \nabla^2W
\right)={\lambda\over r} \left[1\pm{\hbar\over 4}
\sqrt{{2\over\lambda}}\cdot{1\over r^{{1\over 2}}} \right] \qquad
.
\]
We shall follow this point of view and briefly summarize the
connection between the superpotential and the solutions of the
Hamilton-Jacobi equation, an issue fully developed in Reference
\cite{Aai}. Interesting work on the link between 2D classical
integrable systems and SUSY quantum mechanics has also been
performed in \cite{Andr2}. We stress, however, that it is not
equivalent first to solve the HJ equation, define the classical
supercharges, and, then to quantize these latter as to first
quantize the purely bosonic system, solve (\ref{eq:ric}), and
then define the quantum supercharges.

\subsection{Hamiltonian formalism and the Hamilton characteristic
function}

Let the ${\cal N}=2$ classical SUSY Hamiltonian be:
\[
H=\frac{1}{2}\sum_{j=1}^N p_j p_j + \frac{1}{2}\sum_{j=1}^N
\frac{\partial W}{\partial x^j}\frac{\partial W}{\partial x^j} -
i\sum_{j=1}^N\sum_{k=1}^N W_{jk} \theta^j_2 \theta^k_1 \qquad ,
\qquad W_{jk}=\frac{\partial^2 W }{\partial x^j \partial x^k}
\qquad .
\]
The momenta and coordinates in the phase superspace are $p_j,
x^j$, $\theta_1^j, \theta_2^j$, where $\theta_1^j$ and
$\theta_2^j$ are the up and down components of $N$ Grassman
Majorana spinors: $ \theta^j_1 \choose \theta_2^j $, $
\theta_\alpha^j\theta_\beta^k+\theta_\beta^k\theta_\alpha^j=0 $,
$\alpha , \beta=1,2 $.

The Poisson superbrackets of any superfunction on the superspace $
\{F,G\}_P=\frac{\partial F}{\partial p_j} \frac{\partial
G}{\partial x^j} - \frac{\partial F}{\partial x^j} \frac{\partial
G }{\partial p_j}+ i F
\frac{\stackrel{\leftarrow}{\partial}}{\partial \theta_\alpha^j}
\frac{\stackrel{\rightarrow}{\partial}}{\partial \theta_\alpha^j}
G$ are obtained from the Poisson  superstructure defined by the
basic superbrackets:
\[
\{p_j,x^k\}_P=\delta_j^k \hspace{1cm}
\{x^j,x^k\}_P=\{p_j,p_k\}_P=0 \hspace{1cm} \{\theta_\alpha^j,
\theta_\beta^k\}_P=i \delta^{jk} \delta_{\alpha\beta}\qquad .
\]
The classical SUSY charges
\[
Q_1=\sum_{j=1}^N\left(p_j\theta^j_1-\frac{\partial W}{\partial
x^j}\theta^j_2\right)\qquad , \qquad
Q_2=\sum_{j=1}^N\left(p_j\theta^j_2+\frac{\partial W}{\partial
x^j}\theta^j_1\right)
\]
close the classical SUSY algebra: $ \{ Q_1,Q_1 \}_P=\{ Q_2,Q_2
\}_P=2iH$, $\{Q_\alpha , H\}_P=0 $, $  \{ Q_1,Q_2
\}_P=-ip_j\frac{\partial W}{\partial x^j}$.

In the canonical quantization procedure. Poisson superbrackets are
promoted to supercommutators: $[\hat{x}^j , \hat{p}^k ] =
i\delta^{jk}$, $\{\hat{\theta}_\alpha^j , \hat{\theta}_\beta^k \}
= - \delta^{jk}\delta_{\alpha\beta}$. The representation of this
Heisenberg superalgebra by $\hat{p}^j = {1\over i} {\partial \over
\partial x^j}, \hat{x}^j=x^j$, $\hat{\theta}^j_1=\psi_1^j
, \hat{\theta}_2^j=\psi_2^j$, where
$\psi^j_1=\frac{i}{\sqrt{2}}(\psi_+^j+\psi_-^j)=\frac{i}{\sqrt{2}}\gamma^j
\, , \,
\psi^j_2=\frac{1}{\sqrt{2}}(\psi_+^j-\psi_-^j)=-\frac{i}{\sqrt{2}}\gamma^{N+j}
$ are the Majorana $\gamma$-matrices, leads to the quantum
supercharges (\ref{eq:scha}), (\ref{eq:schaa}) and the quantum
superalgebra (\ref{eq:salg}) of Section \S 2.

Setting all the Grassman variables $\theta^j_\alpha$ equal to zero
-the \lq\lq body" of the superspace- , we have a Hamiltonian
dynamical system with Hamiltonian and Hamilton-Jacobi equation:
\[
H=\frac{1}{2}\sum_{j=1}^Np_jp_j+V(x^1,x^2,\cdots ,x^N) \qquad ,
\qquad {\partial S\over \partial t}+H(\frac{\partial S}{\partial
x^1}, \frac{\partial S}{\partial x^2}, \cdots , \frac{\partial S}
{\partial x^n}, x^1,x^2,\cdots ,x^N)=0 \qquad .
\]
There being no explicit dependence on time in $H$, one looks for
solutions of the form $S(t,x^1,x^2,\cdots ,x^N)=W(x^1,x^2,\cdots
,x^N)-i_1t$, and the time-independent Hamilton-Jacobi equation
reads:
\begin{equation}
i_1=\frac{1}{2}\sum_{j=1}^N{\partial W\over\partial x^j}{\partial
W\over\partial x^j}+V(x^1,x^2,\cdots ,x^N) \qquad . \label{eq:thj}
\end{equation}
$W(x^1,x^2,\cdots ,x^N)$ is usually referred to as the Hamilton
characteristic function. Assuming semi-definite positive potential
energy -$U(x^1,x^2, \cdots ,x^N)\geq 0$-, we state the following:

${\it The}$ ${\it superpotential}$ ${\it of}$ ${\it a}$ ${\it
N-dimensional}$ ${\cal N}=2$ ${\it supersymmetric}$ ${\it
dynamical}$ ${\it system}$ ${\it is}$ ${\it a}$ ${\it solution}$
${\it of}$ ${\it the}$ ${\it time-independent}$ ${\it
Hamilton-Jacobi}$ ${\it equation}$ (\ref{eq:thj}) ${\it for}$
$i_1=0$ ${\it and}$ $V(\vec{x})=-U(\vec{x})$.

Therefore, there are as many superpotentials as there are
solutions of the Hamilton-Jacobi equation with zero energy in
minus the potential energy of the body of the supersymmetric
system. More precisely: given a Hamiltonian system with potential
energy $U(x^1,x^2,\cdots ,x^N)$, there are as many ${\cal N}=2$
supersymmetric extensions as there are zero-energy solutions of
the Hamilton-Jacobi equation (\ref{eq:thj}) for $V(x^1,x^2,\cdots
,x^N)=-U(x^1,x^2,\cdots ,x^N)$.

Further understanding of the consequences of this statement is
provided by systems for which the Hamilton-Jacobi equation is
separable. Separability in connection with pseudo-Hermitcity has
been considered in the context of 2D SUSY quantum mechanics in
\cite{Can}. In particular, if $U(x^1,x^2,\cdots ,x^N)=\sum_{j=1}^N
U_j(x^j)$, there are $2^N$ solutions of (\ref{eq:thj}). If there
are no cyclic coordinates,
\[
W^{(a_1,a_2,\cdots ,a_N)} (x^1,x^2,\cdots
,x^N)=(-1)^{a_1}W_1(x^1)+(-1)^{a_2}W_2(x^2)+\cdots
+(-1)^{a_N}W_N(x^N)\qquad ,
\]
where $ a_1,a_2,\cdots ,a_N=0,1$. $N=2$-dimensional systems for
which the Hamilton-Jacobi equation is separable in Cartesian
coordinates are called Type IV Liouville systems; see
\cite{Perelomov}. In this case, changing a global sign in
$W^{(0,0)}$ merely exchanges $\hat{h}^0$ by $\hat{h}^2$ and
$\hat{h}^1_{11}$ by $\hat{h}^1_{22}$: i.e., it is tantamount to
Hodge duality. Choosing $W^{(0,1)}(x^1,x^2)=W_1(x^1)-W_2(x^2)$
instead of $W^{(0,0)}(x^1,x^2)=W_1(x^1)+W_2(x^2)$, one replaces
$W$ by $\tilde{W}$ and the second supersymmetric extension based
on $\tilde{W}$ exhibits a fermionic zero mode if the first
extension has a bosonic zero mode. The other eigenfunctions also
change and the supersymmetric systems are not equivalent.

Even if the Hamilton-Jacobi equation is not separable, one can
still envisage situations where a manifold of solutions is
available. Let us consider a Hamiltonian system with two degrees
of freedom and potential energy:
\[
U(x^1,x^2)=\lambda^2(x^1x^1+x^2x^2)^n-2\lambda\alpha(x^1x^1+x^2x^2)^{{n\over
2}}{\rm cos}\left[n\,\,{\rm arctan}\left\{x^2\over
x^1\right\}\right]+\alpha^2 \qquad ,
\]
where $\lambda$ and $\alpha$ are real physical parameters. It is
not difficult to show, see \cite{Aai1}, that there is a circle of
zero energy solutions of the Hamilton-Jacobi equation with
$V(x^1,x^2)=-U(x^1,x^2)$. If we define
\[
W(x^1,x^2)={\lambda\over n}(x^1x^1+x^2x^2)^{{n\over 2}}{\rm
cos}\left[n\,\,{\rm arctan}\left( {x^2\over x^1}\right)
\right]-\alpha x^1 \qquad ,
\]
\[
{\cal W}(x^1,x^2)={\lambda\over n}(x^1x^1+x^2x^2)^{{n\over 2}}{\rm
sin}\left[n\,\,{\rm arctan}\left( {x^2\over x^1}\right) \right]
-\alpha x^2 \qquad ,
\]
the one-parametric family
\[
\left(\begin{array}{c} W^{(\alpha)}(x^1,x^2) \\ {\cal
W}^{(\alpha)}(x^1,x^2)\end{array}\right)=\left(\begin{array}{cc}
{\rm cos}\alpha & {\rm sin}\alpha \\ -{\rm sin}\alpha & {\rm
cos}\alpha \end{array}\right) \left(\begin{array}{c} W(x^1,x^2)
\\ {\cal W}(x^1,x^2)\end{array}\right)
\]
forms such a circle of solutions. The proof is based on the fact
that $W$ and ${\cal W}$ are harmonic conjugate functions and
satisfy the real analytic Cauchy-Riemann equations $\frac{\partial
W}{\partial x^1}=\frac{\partial {\cal W}}{\partial x^2} $, $
\frac{\partial W}{\partial x^2}=-\frac{\partial {\cal W}}{\partial
x^1} $, a necessary and sufficient condition to build ${\cal N}=4$
supersymmetric extensions in this system.

\subsection{Quantum super Liouville Type I models}
There are other dynamical systems that are Hamilton-Jacobi
separable in two dimensions. We shall focus on systems that are
separable using elliptic coordinates classified by Liouville as
Type I. For a thorough analysis of this kind of ${\cal N}=2$
supersymetric classical system, we refer to \cite{Aai}.
\subsubsection{Classical super Liouville models of Type I}
 Let us consider the map $\xi: {\Bbb R}^2\longrightarrow {\Bbb D}^2$,
where ${\Bbb D}^2$ is an open sub-set of ${\Bbb R}^2$, with
coordinates $(u,v)$, and let $\xi^{-1}: {\Bbb D}^2\longrightarrow
{\Bbb R}^2$ be the inverse map:
\[
(x^1,x^2)=\xi^{-1}(u,v)=\left( \frac{1}{c}uv, \pm \frac{1}{c}
\sqrt{(u^2-c^2)(c^2-v^2)}\right) \qquad , \qquad
\xi(x^1,x^2)=(u,v)
\]
{\small \[
u=\left({\sqrt{(x^1+c)^2+x^2x^2}+\sqrt{(x^1-c)^2+x^2x^2}\over 2}
\right)
 \, , \, v=\left({\sqrt{(x^1+c)^2+x^2x^2}-\sqrt{(x^1-c)^2+x^2x^2}\over 2}\right)
\]}
The $u,v$ variables are the elliptic coordinates of the bosonic
system: $u \in [c,\infty)$, $v \in [-c, c]$ and ${\Bbb D}^2$ is
the closure of the infinite strip: $\bar{\Bbb
D}^2=[c,\infty)\times [-c, c]$. Let us assume the notation $\xi^*$
for the map induced in the functions on ${\Bbb R}^2$; i.e. $\xi^*
U (x^1,x^2)=U(\xi(x^1,x^2))\equiv U(u,v)$. Thus, we shall write
$U$ for $U(x^1,x^2)$ and $\xi^* U$ for $U(u,v)$ and a similar
convention will be used for the functions in the phase and
co-phase spaces.

The Hamilton-Jacobi equation for zero energy and $V=-U$, formula
(\ref{eq:thj}), written in elliptic coordinates, reads:
\begin{equation}
\xi^* U= \displaystyle\frac{u^2-c^2}{u^2-v^2} \,
f(u)+\displaystyle\frac{c^2-v^2}{u^2-v^2} \, g(v) =\frac{1}{2}
\displaystyle\frac{u^2-c^2}{u^2-v^2} \left(\frac{d F}{du}
\right)^2 +\frac{1}{2} \displaystyle\frac{c^2-v^2}{u^2-v^2}
\left(\frac{d G}{dv} \right)^2 \label{eq:ehj}\quad ,
\end{equation}
assuming separability: $\xi^* W=F(u)+G(v)\Rightarrow
\frac{\partial^2 \xi^*W}{\partial u\partial v}=0$. Note that
$f(u),g(v)$ come from the bosonic potential. A complete solution
of (\ref{eq:ehj}) consists of the four combinations of the two
independent one-dimensional problems:
\begin{eqnarray}
F(u)&=&\int du \sqrt{2 f(u)} \qquad , \qquad  G(v)=\int dv
\sqrt{2g(v)}\nonumber \\ \xi^*W^{(a,b)}&=&(-1)^a \int du \sqrt{2
f(u)} +(-1)^b \int dv \sqrt{2g(v)} \quad ,\, a,b=0,1 \quad
\label{eq:sup1}
\end{eqnarray}
The map $\xi^*$ induces a non-Euclidean metric in ${\Bbb
D}^2=(c,\infty)\times (-c,c)$ with metric tensor and Christoffel
symbols:

{\footnotesize
\[ g(u,v)=\left(
\begin{array}{cc} g_{uu}=\displaystyle\frac{u^2-v^2}{u^2-c^2} &
g_{uv}=0 \\ g_{vu}=0 &
g_{vv}=\displaystyle\frac{u^2-v^2}{c^2-v^2}\end{array}
\right)\qquad , \qquad g^{-1}(u,v)=\left( \begin{array}{cc}
g^{uu}=\displaystyle\frac{u^2-c^2}{u^2-v^2} & g^{uv}=0 \\ g^{vu}=0
& g^{vv}=\displaystyle\frac{c^2-v^2}{u^2-v^2}
\end{array} \right)
\]}

{\small
\[
\begin{array}{lclcl}
\Gamma_{uu}^u=\displaystyle\frac{-u (c^2-v^2)}{(u^2-v^2)(u^2-c^2)}
& ,& \Gamma_{vv}^v=\displaystyle\frac{v
(u^2-c^2)}{(u^2-v^2)(c^2-v^2)} & ,&
\Gamma_{uv}^u=\Gamma_{vu}^u=\displaystyle\frac{-v}{u^2-v^2}
\\[0.4cm] \Gamma_{uu}^v=\displaystyle\frac{v
(c^2-v^2)}{(u^2-v^2)(u^2-c^2)} & ,&
\Gamma_{vv}^u=\displaystyle\frac{-u (u^2-c^2)}{(u^2-v^2)(c^2-v^2)}
&,& \Gamma_{uv}^v=\Gamma_{vu}^v=\displaystyle\frac{u }{u^2-v^2}
\end{array}
\]}

\noindent Besides the bosonic (even Grassman) variables $u$, $v$,
there are also fermionic (odd Grassman) Majorana spinors
$\vartheta^u_\alpha$ , $\vartheta^v_\alpha$ in the system. We
choose the zweibein
\[
g^{uu}(u,v)=\sum_{j=1}^2e^u_j(u,v)e^u_j(u,v) \qquad , \qquad
g^{vv}(u,v)=\sum_{j=1}^2e^v_j(u,v)e^v_j(u,v)
\]
in the form:
\[
e^u_1(u,v)=\left(u^2-c^2\over u^2-v^2\right)^{{1\over 2}} \qquad ,
\qquad e^v_2(u,v)=\left(c^2-v^2\over u^2-v^2\right)^{{1\over
2}}\qquad .
\]
Curved and flat Grassman variables are related as: $
\vartheta^u_\alpha(u,v)=e^u_1(u,v)\theta^1_\alpha $, $
\vartheta^v_\alpha(u,v)=e^v_1(u,v)\theta^1_\alpha $.

A supersymmetric two-dimensional mechanical system is a super-
Liouville model of Type I if the Lagrangian is of the form
$\xi^*L=\xi^*L_B+\xi^*L_F+\xi^*L_{BF}$, with:
\begin{eqnarray*}
\xi^*L_B&=&{1\over 2}g_{uu}(u,v)\dot{u}\dot{u}+{1\over
2}g_{vv}(u,v)\dot{v}\dot{v}-{1\over 2}g^{uu}(u,v)\left({dF\over
du}\right)^2-{1\over 2}g^{vv}(u,v)\left({dG\over dv}\right)^2\\
\xi^*L_F&=&-\frac{i}{2} g_{uu}(u,v) \vartheta_\alpha^u D_t
\vartheta_\alpha^u- \frac{i}{2} g_{vv}(u,v)\vartheta_\alpha^v D_t
\vartheta_\alpha^v \\ \xi^*L_{BF}^I & =& -i \left[ \frac{d^2
F}{du^2}-\Gamma^u_{uu} \frac{dF}{du} -\Gamma^v_{uu}\frac{dG}{dv}
\right] \vartheta_2^u \vartheta_1^u  -i \left[ \frac{d^2 G}{dv^2}
- \Gamma^u_{vv} \frac{dF}{du} -\Gamma^v_{vv} \frac{dG}{dv}
\right] \vartheta_2^v \vartheta_1^v+
\\ & & +i \left[ \Gamma^u_{uv}
\frac{dF}{du}-\Gamma^v_{uv} \frac{dG}{dv}  \right] (\vartheta_2^v
\vartheta_1^u+\vartheta_2^u \vartheta_1^v) \qquad .
\end{eqnarray*}
The fermionic kinetic energy is encoded in $\xi^*L_F$, where the
covariant derivatives are defined as:
\begin{eqnarray*}
D_t\vartheta^u_\alpha &=& \dot{\vartheta}^u_\alpha
+\Gamma^u_{uu}\dot{u}\vartheta^u_\alpha +
\Gamma^u_{uv}\dot{u}\vartheta^v_\alpha
+\Gamma^u_{vu}\dot{v}\vartheta^u_\alpha
+\Gamma^u_{vv}\dot{v}\vartheta^u_\alpha \\
D_t\vartheta^v_\alpha&=& \dot{\vartheta}^v_\alpha
+\Gamma^v_{uu}\dot{u}\vartheta^u_\alpha +
\Gamma^v_{uv}\dot{u}\vartheta^v_\alpha
+\Gamma^v_{vu}\dot{v}\vartheta^u_\alpha
+\Gamma^v_{vv}\dot{v}\vartheta^v_\alpha \qquad .
\end{eqnarray*}
The Yukawa terms governing the Bose-Fermi interactions are
prescribed in $\xi^*L_{BF}$. The generalized momenta of the
supersymmetric system and the supercharges are:
\begin{eqnarray*}
P_u&=& g_{uu}(u,v)\left(\dot{u}-{i\over
2}\vartheta^u_\alpha\Gamma^u_{uv}\vartheta^v_\alpha\right)-{i\over
2}g_{vv}(u,v)\vartheta^v_\alpha\Gamma^v_{uu}\vartheta^u_\alpha \\
P_v&=& g_{vv}(u,v)\left(\dot{v}-{i\over
2}\vartheta^v_\alpha\Gamma^v_{vu}\vartheta^u_\alpha\right)-{i\over
2}g_{uu}(u,v)\vartheta^u_\alpha\Gamma^u_{vv}\vartheta^v_\alpha
\quad , \\ \xi^*Q_1^{(a,b)}&=&P_u\vartheta^u_1+{i\over
2}\left(g_{uu}\vartheta^u_\alpha\Gamma^u_{uv}\vartheta^v_\alpha+
g_{vv}\vartheta^v_\alpha\Gamma^v_{uu}\vartheta^u_\alpha\right)\vartheta^u_1
-(-1)^a{dF\over du}\vartheta^u_2 \\&&+ P_v\vartheta^v_1+{i\over
2}\left(g_{vv}\vartheta^v_\alpha\Gamma^v_{vu}\vartheta^u_\alpha+
g_{uu}\vartheta^u_\alpha\Gamma^u_{vv}\vartheta^v_\alpha\vartheta^u_1\right)\vartheta^v_1
-(-1)^b{dG\over dv}\vartheta^v_2 \\
\xi^*Q_2^{(a,b)}&=&P_u\vartheta^u_2+{i\over
2}\left(g_{uu}\vartheta^u_\alpha\Gamma^u_{uv}\vartheta^v_\alpha+
g_{vv}\vartheta^v_\alpha\Gamma^v_{uu}\vartheta^u_\alpha\right)\vartheta^u_2
+(-1)^a{dF\over du}\vartheta^u_1 \\&&+ P_v\vartheta^v_2+{i\over
2}\left(g_{vv}\vartheta^v_\alpha\Gamma^v_{vu}\vartheta^u_\alpha+
g_{uu}\vartheta^u_\alpha\Gamma^u_{vv}\vartheta^v_\alpha\vartheta^u_1\right)\vartheta^u_2
+(-1)^b{dG\over dv}\vartheta^v_1 \qquad .
\end{eqnarray*}

\subsubsection{Quantum supercharges and Hamiltonian}
Passing to Majorana-Weyl spinors,
$\vartheta^{u,v}_{+}={1\over{\sqrt 2}}(\vartheta^{u,v}_2-i
\vartheta^{u,v}_1)$, $\vartheta^{u,v}_{-}=-{1\over{\sqrt
2}}(\vartheta^{u,v}_2+ i \vartheta^{u,v}_1)$, the fermionic
quantization rules lead us to the Fermi operators in non-Euclidean
space: $\psi_{\pm}^{u}(u,v)=e^u_1(u,v)\psi^1_\pm \, , \,
\psi_\pm^v(u,v)=e^v_2(u,v)\psi_\pm^2$. Setting e.g. $a=b=0$, and
also quantizing the generalized momenta, $\hat{P}_u={1\over
i}{\partial\over\partial u}$, $\hat{P}_v={1\over
i}{\partial\over\partial v}$, we obtain the quantum supercharges:
\[
\xi^*\hat{Q}_\pm=- i  \psi_\pm^u \nabla_u^\mp - i {u\over c^2-v^2}
\psi_\pm^u \psi_\pm^v \psi_\mp^v -i\psi_\pm^v \nabla_v^\mp + i
{v\over u^2-c^2} \psi_\pm^v \psi_\pm^u \psi_\mp^u \qquad ;
\]
or, in matrix form:

{\footnotesize
\begin{equation} \xi^*\hat{Q}_+=-
i\left(\begin{array}{cccc} 0 & 0 & 0 & 0 \\ e^u_1\nabla_u^- & 0 &
0 & 0
\\ e^v_2\nabla_v^- & 0 & 0 & 0
\\ 0 & -e^v_2 \left( \nabla_v^- -{v\over u^2-v^2} \right)
 & e^u_1 \left( \nabla_u^- +{u\over u^2-v^2} \right)  & 0
\end{array}\right)\qquad , \qquad \nabla_u^\mp={\partial\over\partial u}\mp {dF\over du}
\qquad , \label{eq:scel1}
\end{equation}}

{\footnotesize
\begin{equation}
\xi^*\hat{Q}_-=-i\left(\begin{array}{cccc} 0 & e^u_1 \left(
\nabla_u^+ +{u\over u^2-v^2} \right) & e^v_2 \left( \nabla_v^+
-{v\over u^2-v^2} \right) & 0
\\ 0 & 0 & 0 & -e^v_2\nabla_v^+
\\ 0 & 0 & 0 & e^u_1\nabla_u^+
\\ 0 & 0 & 0 & 0
\end{array}\right)\quad , \qquad \nabla_v^\mp={\partial\over\partial v}\mp {dG\over
dv}\qquad . \label{scel2}
\end{equation}}

In order to make clear how separability and supersymmetry are
entangled, it is convenient to write the different pieces of the
quantum Hamiltonian , $\xi^*\hat{H}={1\over
2}\{\xi^*\hat{Q}_+,\xi^*\hat{Q}_-\}$,
\begin{equation}
\xi^*{\hat H} = {1\over 2 (u^2-v^2)} \left(
\begin{array}{ccc} \xi^*{\hat h}^{(0)}({\partial\over \partial u},{\partial\over \partial v}
,u,v) & 0 & 0  \\ 0  & \xi^*{\hat h}^{(1)}({\partial\over
\partial u},{\partial\over \partial v} ,u,v) & 0
\\ 0 & 0 & \xi^*{\hat h}^{(2)}({\partial\over \partial u},{\partial\over \partial v}
,u,v)
\end{array} \right)\qquad ,
\end{equation}
separately. On the subspaces ${\cal H}_0$ and ${\cal H}_2$, the
differential operator $\xi^*\hat{H}$ splits into the following
structure:
\begin{itemize}
\item $\xi^*\hat{h}^{(0)}({\partial\over\partial u},{\partial\over\partial v}
,u,v)=\hat{j}^{(0)}({\partial\over\partial
u},u)+\hat{k}^{(0)}({\partial\over\partial v} ,v)$.
\[
\hat{j}^{(0)}({\partial\over\partial
u},u)=(u^2-c^2)\left[-{\partial^2\over\partial u^2}-{u\over
u^2-c^2}{\partial\over\partial u}+\left({dF\over du}\right)^2+
{d^2F\over du^2}+{u\over u^2-c^2}{dF\over du}\right]
\]
\[
\hat{k}^{(0)}({\partial\over\partial
v},v)=(c^2-v^2)\left[-{\partial^2\over\partial v^2}+{v\over
c^2-v^2}{\partial\over\partial v}+\left({dG\over dv}\right)^2+
{d^2G\over dv^2}-{v\over c^2-v^2}{dG\over dv}\right]
\]

\item $\xi^*\hat{h}^{(2)}({\partial\over
\partial u},{\partial\over\partial v}
,u,v)=\hat{j}^{(2)}({\partial\over\partial
u},u)+\hat{k}^{(2)}({\partial\over \partial v} ,v)$.
\[
\hat{j}^{(2)}({\partial\over\partial
u},u)=(u^2-c^2)\left[-{\partial^2\over\partial u^2}-{u\over
u^2-c^2}{\partial\over\partial u}+\left({dF\over du}\right)^2-
{d^2F\over du^2}-{u\over u^2-c^2}{dF\over du}\right]
\]
\[
\hat{k}^{(2)}({\partial\over\partial
v},v)=(c^2-v^2)\left[-{\partial^2\over\partial v^2}+{v\over
c^2-v^2}{\partial\over\partial v}+\left({dG\over dv}\right)^2-
{d^2G\over dv^2}+{v\over c^2-v^2}{dG\over dv}\right] \qquad .
\]
\end{itemize}
Therefore, we conclude that in the bosonic sectors the dynamical
problem is separable in the $u$ and $v$ variables.

Things, however, become more involved in the fermionic sectors. We
write the Hamiltonian acting on ${\cal H}_1$ as follows:
{\footnotesize\[ \xi^*\hat{h}^{(1)}({\partial\over\partial
u},{\partial\over\partial v},u,v)=\left(\begin{array}{cc}
l^{(1)}_+({\partial\over\partial
u},u)+f^{(1)}_+({\partial\over\partial v},v)+g^{(1)}_+(u,v) &
t^{(1)}_+(u,v) \\ & \\ t^{(1)}_-(u,v) &
l^{(1)}_-({\partial\over\partial
u},u)+f^{(1)}_-({\partial\over\partial v},v)+g^{(1)}_-(u,v)
\end{array}\right) \qquad .
\]}
Here,
\[
l^{(1)}_\pm ({\partial\over\partial u},u)=(u^2-c^2) \left[
-{\partial^2 \over \partial u^2} - {u\over u^2-c^2} {\partial
\over \partial u} + \left( { d F \over d u} \right)^2\mp {d^2 F
\over d u^2}\right]\qquad ,
\]
{\footnotesize\[ f^{(1)}_\pm ({\partial\over\partial
u},u)=(c^2-v^2) \left[ -{\partial^2 \over \partial v^2} + {v\over
c^2-v^2} {\partial \over
\partial v} + \left( { d G \over d v} \right) ^2\pm {d^2 G \over d
v^2}\right]\qquad ,
\]}
{\footnotesize\[ g^{(1)}_\pm (u,v)={(u^2+v^2-2c^2)\over
u^2-v^2}\left[\pm u{dF\over du}\mp v{dG\over dv} \right]\quad ,
\quad t^{(1)}_\pm (u,v)={ \sqrt{(u^2-c^2)(c^2-v^2)} \over
(u^2-v^2)^2} \left( v {d F\over d u}+  u {d G\over d v}
\right)\qquad .
\]}

\noindent The variables $u$ and $v$ are mixed in
$\xi^*\hat{h}^{(1)}$. It seems that supersymmetry breaks down
separability. Nevertheless, the non-null spectrum of
$\xi^*\hat{h}^{(1)}$ is given by the non-null spectra of
$\xi^*\hat{h}^{(0)}$ and $\xi^*\hat{h}^{(2)}$, operators with
separable spectral problems.

\section{Two examples in two dimensions}

\subsection{The Planar anisotropic harmonic oscillator}
This is a Type IV Liouville model. If $a_1,a_2=0,1$, the
potential, superpotentials and supercharges are:
\[
U(x_1,x_2)={k_1\over 2}x_1^2+{k_2\over 2}x_2^2 \qquad , \qquad
W^{(a_1,a_2)}(x_1,x_2)={(-1)^{a_1}\over
2}\sqrt{k_1m_1}x_1^2+{(-1)^{a_2}\over 2}\sqrt{k_2m_2}x_2^2
\]
\[
\hat{Q}_+^{(a_1,a_2)}=i\sqrt{2}{\footnotesize\left(\begin{array}{cccc}
0 & 0 & 0 & 0
\\ \hat{q}_1^{(a_1)} &
0 & 0 & 0 \\ \hat{q}_2^{(a_2)} & 0 & 0 & 0
\\ 0 & -\hat{q}_2^{(a_2)} &
\hat{q}_1^{(a_1)} & 0
\end{array}\right)}\,\,\, ; \,\,\,
\hat{Q}_-^{(a_1,a_2)}=i\sqrt{2}{\footnotesize
\left(\begin{array}{cccc} 0 & \hat{q}_1^{(a_1)^\dagger} &
\hat{q}_2^{(a_2)^\dagger} & 0
\\ 0 & 0 & 0 & -\hat{q}_2^{(a_2)^\dagger}
\\ 0 & 0 & 0 & \hat{q}_1^{(a_1)^\dagger}
\\ 0 & 0 & 0 & 0
\end{array}\right)}
\]
\[
\hat{q}_1^{(a_1)}={1\over\sqrt{2}}{\partial\over\partial
x_1}-(-1)^{a_1}\sqrt{{k_1m_1\over 2}}x_1 \qquad , \qquad
\hat{q}_2^{(a_2)}={1\over\sqrt{2}}{\partial\over\partial
x_2}-(-1)^{a_2}\sqrt{{k_2m_2\over 2}}x_2 \qquad .
\]
From the annihilation operators
\[
\hat{A}_1={1\over\sqrt{2m_1}}{\partial\over\partial
x_1}+\sqrt{k_1\over 2}x_1 \qquad , \qquad
\hat{A}_2={1\over\sqrt{2m_2}}{\partial\over\partial
x_2}+\sqrt{k_2\over 2}x_2 \qquad ,
\]
their adjoints, and the natural frequencies
$\omega_1=\sqrt{{k_1\over m_1}}$, $\omega_2=\sqrt{{k_2\over
m_2}}$ one obtains the Hamiltonian:
\[
\hat{h}^{(0)}=\sum_{j=1}^2\omega_j\left(\hat{A}_j^\dagger\hat{A}_j+{1\over
2}(1+(-1)^{a_j})\right) \qquad , \qquad
\hat{h}^{(2)}=\sum_{j=1}^2\omega_j\left(\hat{A}_j^\dagger\hat{A}_j+{1\over
2}(1-(-1)^{a_j})\right)\qquad ,
\]
\[
\hat{h}^{(1)}={\footnotesize\left(\begin{array}{cc}
\sum_{j=1}^2\omega_j\left(\hat{A}_j^\dagger\hat{A}_j+{1\over
2}\right)-{(-1)^{a_1}\over 2}\omega_1+{(-1)^{a_2}\over 2}\omega_2 & 0 \\
0 & \sum_{j=1}^2\omega_j\left(\hat{A}_j^\dagger\hat{A}_j+{1\over
2}\right)+{(-1)^{a_1}\over 2}\omega_1-{(-1)^{a_2}\over 2}\omega_2
\end{array}\right)}\qquad .
\]
The Fock space basis
\[
\hat{A}_1|0,0\rangle=\hat{A}_2|0,0\rangle =0 \qquad , \qquad
|n_1,n_2\rangle={1\over\sqrt{n_1!n_2!}}
(\hat{A}_1^\dagger)^{n_1}(\hat{A}_2^\dagger)^{n_2}|0,0\rangle
\]
provides the eigenfunctions:
$\hat{A}_1^\dagger\hat{A}_1|n_1,n_2\rangle=n_1|n_1,n_2\rangle $, $
\hat{A}_2^\dagger\hat{A}_2|n_1,n_2\rangle=n_2|n_1,n_2\rangle $.
Thus,
\[
{\rm Spec}\,\,\hat{h}^{(0)}=\sqcup_{j=1}^2\omega_j[n_j+{1\over
2}(1+(-1)^{a_j})] \qquad , \qquad {\rm
Spec}\,\,\hat{h}^{(2)}=\sqcup_{j=1}^2\omega_j[n_j+{1\over
2}(1-(-1)^{a_j})]
\]
\[
{\rm Spec}\,\,\hat{h}^{(1)}=\sqcup_{j=1}^2\omega_j(n_j+{1\over
2})\mp {(-1)^{a_1}\over 2}\omega_1\pm {(-1)^{a_2}\over 2}\omega_2
\qquad .
\]
The ground state
\[
\Psi(x_1,x_2)=\langle x_1,x_2|0,0\rangle ={\rm exp}[-{1\over
2}\sum_{j=1}^2 \omega_jm_jx_j^2 ]
\]
belongs to: (a) ${\cal H}_0$, if $a_1=a_2=1$, (b) ${\cal H}_2$, if
$a_1=a_2=0$, (c) ${\cal H}_1$, if $a_1\neq a_2$. For the
$a_1=a_2=1$ case, the SUSY partner states - all of them with
energy $E=n_1\omega_1+n_2\omega_2$ - are:
\[
\begin{array}{ccccc}
&&{\tiny \left(\begin{array}{c} |n_1,n_2\rangle \\ 0 \\
0 \\ 0
\end{array}\right)}&& \\
\hspace{3cm}\hat{Q}_+ & \swarrow&&\searrow& \hspace{-2.cm} \hat{Q}_+\\
{\tiny \left(\begin{array}{c} 0 \\ |n_1-1,n_2\rangle \\ 0 \\ 0
\end{array}\right)}&&&&{\tiny \left(\begin{array}{c} 0 \\ 0 \\ |n_1,n_2-1\rangle \\ 0
\end{array}\right)} \\
\hspace{3cm}\hat{Q}_+&\searrow&&\swarrow& \hspace{-2.cm} \hat{Q}_+\\
 &&{\tiny \left(\begin{array}{c} 0 \\ 0 \\ 0 \\
|n_1-1,n_2-1\rangle
\end{array}\right)}&&
\end{array}
\]
Other choices for $a_1$ and $a_2$ require permutations between
the vertices of the rhombus.

\subsection{Two Newtonian centers of force on a plane}

Let us start with the energy potential for the problem of two
attractive centers of force with non-equal strengths (see Figure
1(a)):
\[
U(x_1,x_2) = - \left( {\alpha_1 \over r_1 }+{\alpha_2 \over
r_2}\right)\qquad , \qquad  0<\alpha_2<\alpha_1 \qquad ,
\]
\noindent\begin{figure}[htbp] \centerline{
 \hspace{1cm}\epsfig{file=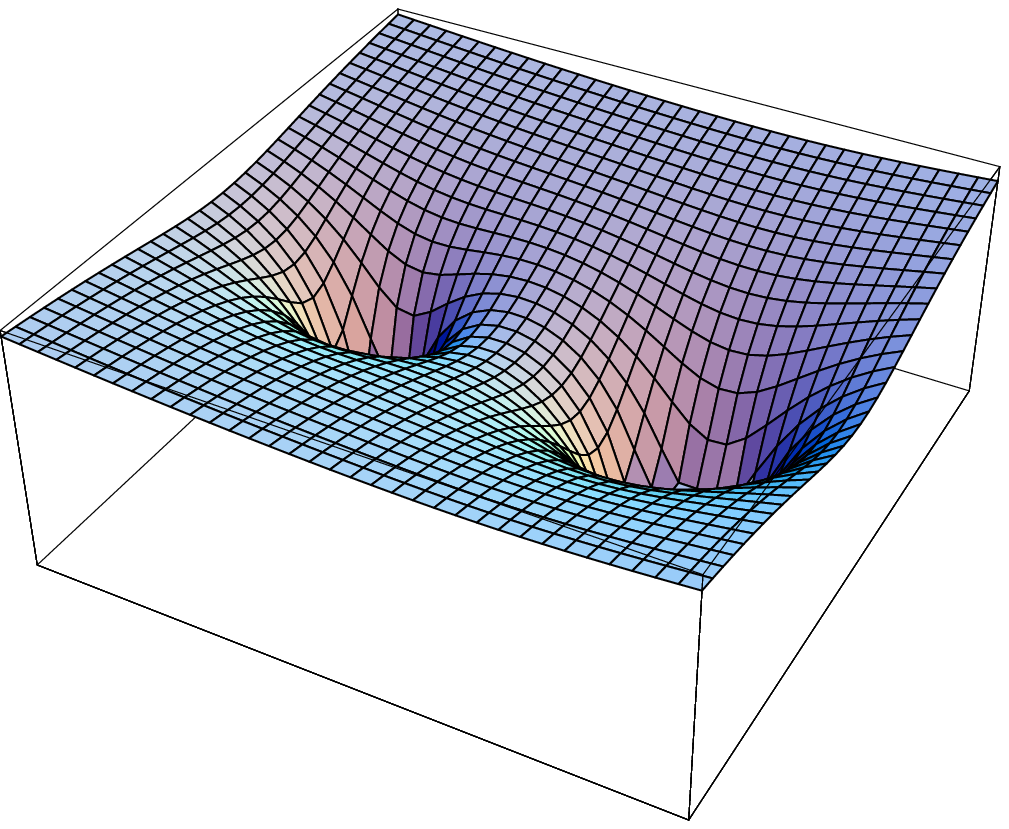,height=4cm}
\hspace{1cm} \epsfig{file=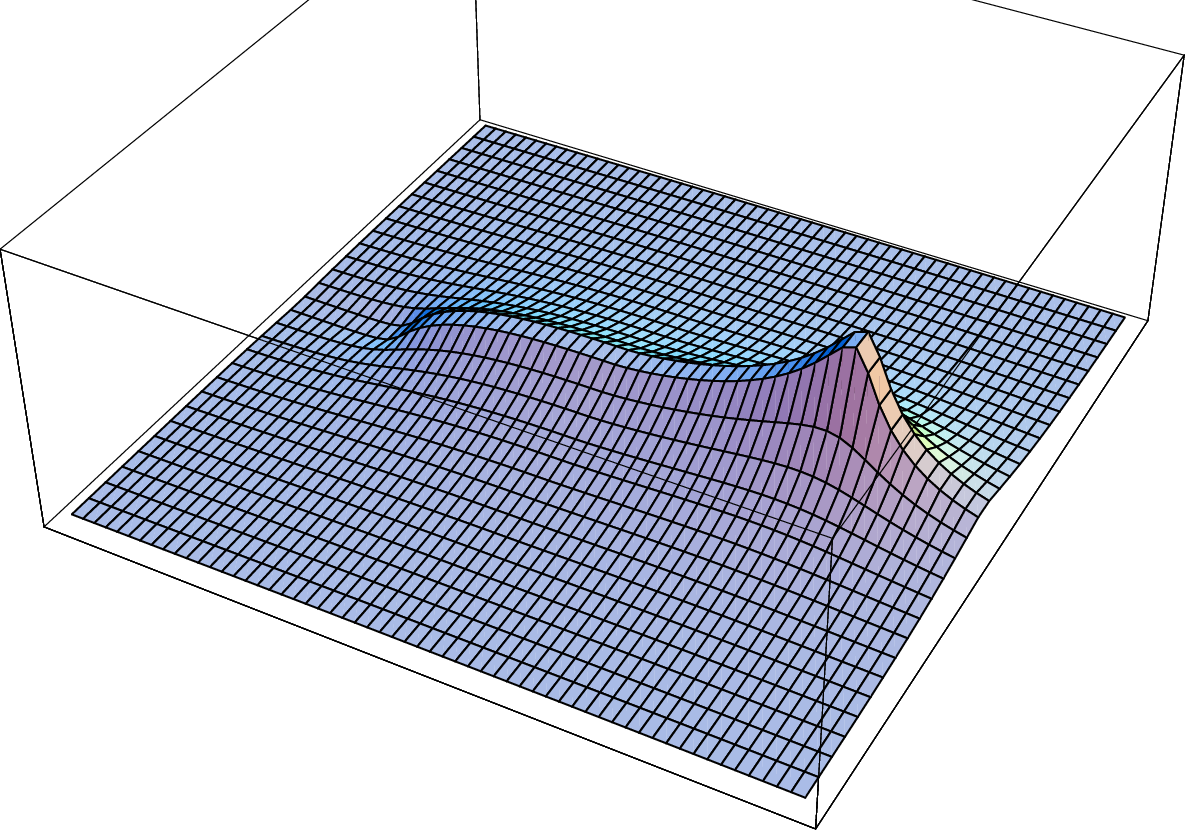,height=4cm} }
\caption{\small a) $U(x_1,x_2)$, $c=1$, $\alpha_1=2$,
$\alpha_2=1$.  b) ${\rm exp} \left(W(x_1,x_2) \right) $. }
\end{figure}

The distances from the centers are appropriately given in terms of
the elliptic coordinates: $u = {1\over 2} (r_1 + r_2)$, $ v =
{1\over 2} (r_2 - r_1)$, $r_1 = \sqrt{(x_1- c)^2 + x_2^2} \,, r_2
= \sqrt{ (x_1 + c )^2 + x_2^2}$. The Hamiltonian in elliptic
coordinates,
\[
\xi^* H = {1\over 2}{1\over u^2 -v^2} \left[ (u^2 -c^2) p_u^2 +
(c^2 - v^2) p_v^2 \right] + { k_+ u - k_- v \over u^2 -v^2} \qquad
,
\]
depends on the coupling constants $ k_\pm = \alpha_2 \pm \alpha_1$
and shows that we are dealing with a type I Liouville system. The
ansatz $S = - i_1 t + F[u;i_1,i_2] + G[v;i_1,i_2]$ leads to the
$i_1=i_2=0$ Hamilton-Jacobi equation:
\begin{equation}
{k_+u - k_- v\over u^2-v^2}={1\over 2}g^{uu}\left({dF\over d
u}\right)^2+{1\over 2}g^{vv}\left({dG\over d v}\right)^2\qquad .
\label{eq:tchj}
\end{equation}
Note that in this case the potential energy is semi-definite
negative and, to find real solutions, we do not replace $U$ by
$-U$ in the Hamilton-Jacobi equation. Therefore, the solution of
(\ref{eq:tchj}) in terms of the elliptic and complete elliptic
integrals of the first and second kind, \cite{Abr},
\[
F(u) =- 2  \sqrt{ k_+ c} \left[ {\rm F}\left( {\rm sin}^{-1}
\sqrt{u-c \over u},{1\over 2} \right) - 2 {\rm E}\left( {\rm
sin}^{-1} \sqrt{u-c \over u},{1\over 2} \right) +  \sqrt{ 2
(u^2-c^2) \over u c} \right]\qquad ,
\]
\[
G(v) =  \left\{ \begin{array}{cc}  2  \sqrt{k_- c} \left[ 2 {\rm
E}\left({\rm sin}^{-1}\sqrt{ 2 v\over v-c},{1\over 2}\right)-{\rm
F}\left({\rm sin}^{-1} \sqrt{2 v\over v-c}, {1\over 2}\right) -
\sqrt{ 2 v (v+c) \over c (v-c)} \right]
 &  -c < v \leq 0 \\ & \\ 2 i \sqrt{k_- c} \left[ 2 {\rm
E}\left({\rm sin}^{-1}\sqrt{c-v\over c},{1\over 2}\right)- 2 {\rm
E} [1/2]+ {\rm F}\left({\rm sin}^{-1} \sqrt{c-v\over c}, {1\over
2}\right) - {\rm K}[1/2] \right] & 0 \leq v < c
\end{array} \right. \qquad ,
\]
provide the superpotentials
\[
\xi^*W^{(a,b)}(x_1,x_2)=(-1)^aF(u)+(-1)^bG(v)
\]
for two ${\it repulsive}$ Newtonian centers. Nevertheless, the
Laplacian of the superpotential - given by the terms
\[
{dF\over d u}=-\sqrt{{2k_+ u\over u^2-c^2}} \qquad , \qquad
{dG\over d v}=-\sqrt{-2k_-  v\over c^2-v^2}
\]
\[
{d^2F\over d u^2}=-{1\over 2}{u^2+c^2\over u(u^2-c^2)}{dF\over d
u}\qquad , \qquad {d^2G\over d v^2}={1\over 2}{c^2+v^2\over
v(c^2-v^2)}{dG\over d v}
\]
coming from the quantization of the Yukawa couplings- induce
attractive forces in the supersymmetric extension of two
repulsive Newtonian centers and there is hope of finding
normalizable eigenstates.

In fact, choosing $a=b=0$ we obtain the zero-energy wave function
in the Bose/Bose sector, $\hat{Q}_+ \Psi_0(x_1,x_2)=0$ :
{\footnotesize\[ i\left(\begin{array}{cccc} 0 & 0 & 0 & 0 \\
e^u_1({\partial\over\partial u}-{dF\over du}) & 0 & 0 & 0
\\ e^v_2({\partial\over\partial v}-{dG\over dv}) & 0 & 0 & 0
\\ 0 & -e^v_2({\partial\over\partial v}-{dG\over dv}-{ v\over u^2-v^2})
 & e^u_1({\partial\over\partial u}
-{dF\over du}+{u \over u^2-v^2})  & 0
\end{array}\right)
\left(
\begin{array}{c} \psi_0(u,v)
\\ 0 \\ 0 \\ 0
\end{array} \right)= 0
\]}
if
\[
e^u_1{\partial\log \psi_0\over\partial u}(u,v)\vec{e}_1+
e^v_2{\partial\log\psi_0\over\partial v}(u,v)\vec{e}_2 =
e^u_1{dF\over du}\vec{e}_1+e^v_2{dG\over dv}\vec{e}_2 \qquad .
\]
Therefore,
\begin{equation}
\psi_0(u,v)=\xi^*\Psi_0^{(0)}(x_1,x_2)=C{\rm exp}[F(u)+ G(v)]
\qquad ,  \label{eq:tcgs}
\end{equation}
which is normalizable, see Figure 2, is the ground state of the
${\cal N}=2$ supersymmetric particle, even though the particle's
\lq\lq body" is repelled by two centers.

\noindent\begin{figure}[htbp] \centerline{
 \hspace{1cm} \epsfig{file=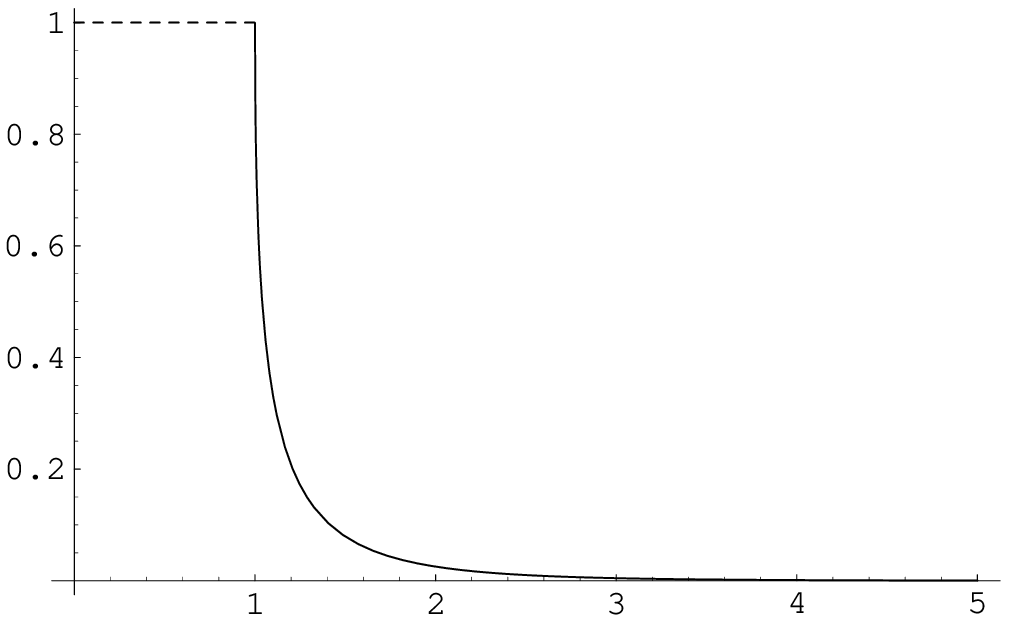,height=3cm}
 \hspace{1cm} \epsfig{file=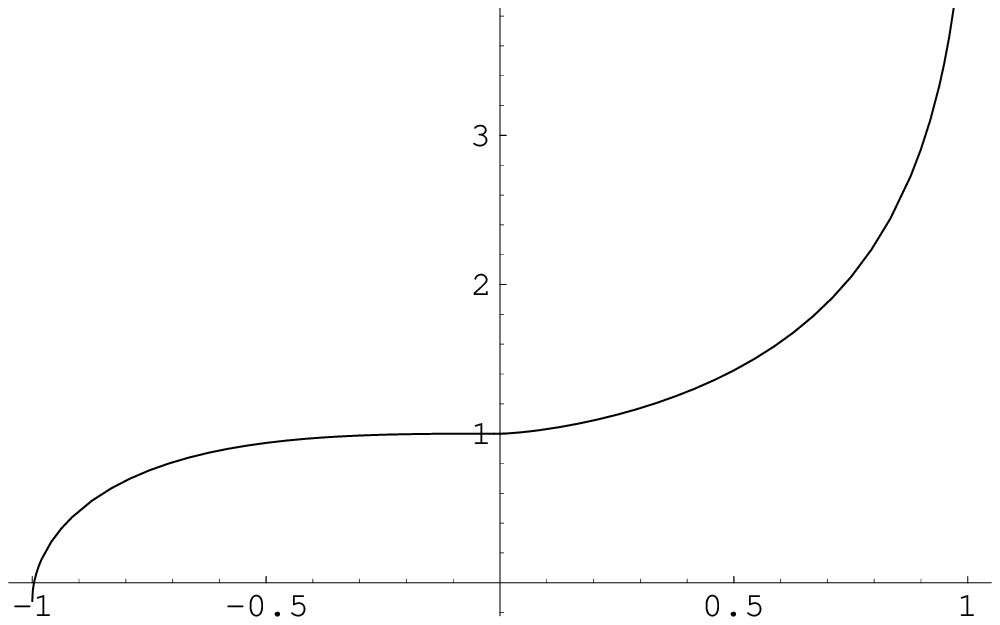,height=3cm}} \caption{\small Plot of ${\rm
exp}[F(u)]$ and ${\rm exp}[G(v)]$ as a function of u and v
respectively.}
\end{figure}
\bigskip

Figure 1(b) shows a plot of the (\ref{eq:tcgs}) wave function in
Cartesian coordinates. It is amusing to check how well it fits in
with the expected behaviour of a quantum particle in a potential
well with two Newtonian holes; see, e.g. , \cite{Pauling} to find
an approximate wave function for the ground state of the
molecule-ion of hydrogen. The reason is that the effective quantum
potential in the ${\cal H}_0$ sub-space
\[
\hat{V}^{(0)}(x_1,x_2)={1\over 2}
\left(\Psi_0^{(0)}\right)^{-1}(x_1,x_2)\nabla^2\Psi_0^{(0)}(x_1,x_2)
\]
{\small \[ \xi^*\hat{V}^{(0)}(x_1,x_2)={1\over 2}
\frac{\psi_0^{-1}(u,v)}{u^2-v^2}\left\{(u^2-c^2)
\left({\partial\over\partial u}+{u\over u^2-c^2}\right)
\frac{\partial\psi_0}{\partial u}(u,v)+(c^2-v^2)\left(
{\partial\over\partial v}- {v\over c^2-v^2}\right)
\frac{\partial\psi_0}{\partial v}(u,v) \right\}
\]}
is attractive towards the two centers.
\section{Summary}
Interactions in supersymmetric classical or quantum mechanics are
prescribed by superpotentials. In this paper we have dealt with
the following inverse problem: Given a Hamiltonian system, is
there a superpotential from which forces are derived? If so, a
supersymetric extension of this particular physical system is
possible. We have encountered a two-fold way to meeting
ambiguities in answering this question.

\begin{enumerate}
\item First, the outcome depends on the framework. For classical
systems, superpotentials are solutions of zero-energy
time-independent Hamilton-Jacobi equations. In the quantum domain
superpotentials solve Ricatti-like PDE's. Moreover, canonical
quantization and supersymmetric extension do not commute: the
supersymmetric extension of - e.g., the quantum Coulomb problem-
differs fom the quantization of the classical supersymmetric
Coulomb system.

\item In dimensions higher than one, superpotentials are far from
unique. For instance, in Hamilton-Jacobi separable systems there
are $2^N$ different superpotentials leading to supersymmetric
systems with the same \lq\lq body" dynamics. The ground states can
be easily found in this kind of system because one needs to solve
only first-order ODE's: one per each variable in which the
dynamics separates.
\end{enumerate}

\end{document}